\documentclass{article}
\PassOptionsToPackage{numbers,sort&compress}{natbib}
\usepackage[preprint]{neurips_2026}
\usepackage[utf8]{inputenc}
\usepackage[T1]{fontenc}
\usepackage{amsmath,amssymb,amsfonts}
\usepackage{algorithmic}
\usepackage{graphicx}
\graphicspath{{figures/}}
\usepackage{float}
\usepackage{subcaption}
\usepackage{textcomp}
\usepackage{xcolor}
\usepackage[hidelinks]{hyperref}
\usepackage{url}
\usepackage{booktabs}
\usepackage{nicefrac}
\usepackage{microtype}
\usepackage{listings}
\lstdefinestyle{tuicode}{
  basicstyle=\ttfamily\scriptsize,
  keywordstyle=\bfseries,
  commentstyle=\itshape\color{gray},
  columns=fullflexible,
  showstringspaces=false,
  breaklines=true,
  numbers=none,
  frame=none,
  aboveskip=2pt,
  belowskip=2pt,
  xleftmargin=2pt,
}
\def\BibTeX{{\rm B\kern-.05em{\sc i\kern-.025em b}\kern-.08em
    T\kern-.1667em\lower.7ex\hbox{E}\kern-.125emX}}

\begin{document}

\title{\texorpdfstring{%
  \href{https://github.com/tui-testing/tuibot}{%
    \raisebox{-0.30\height}{\includegraphics[height=0.50in]{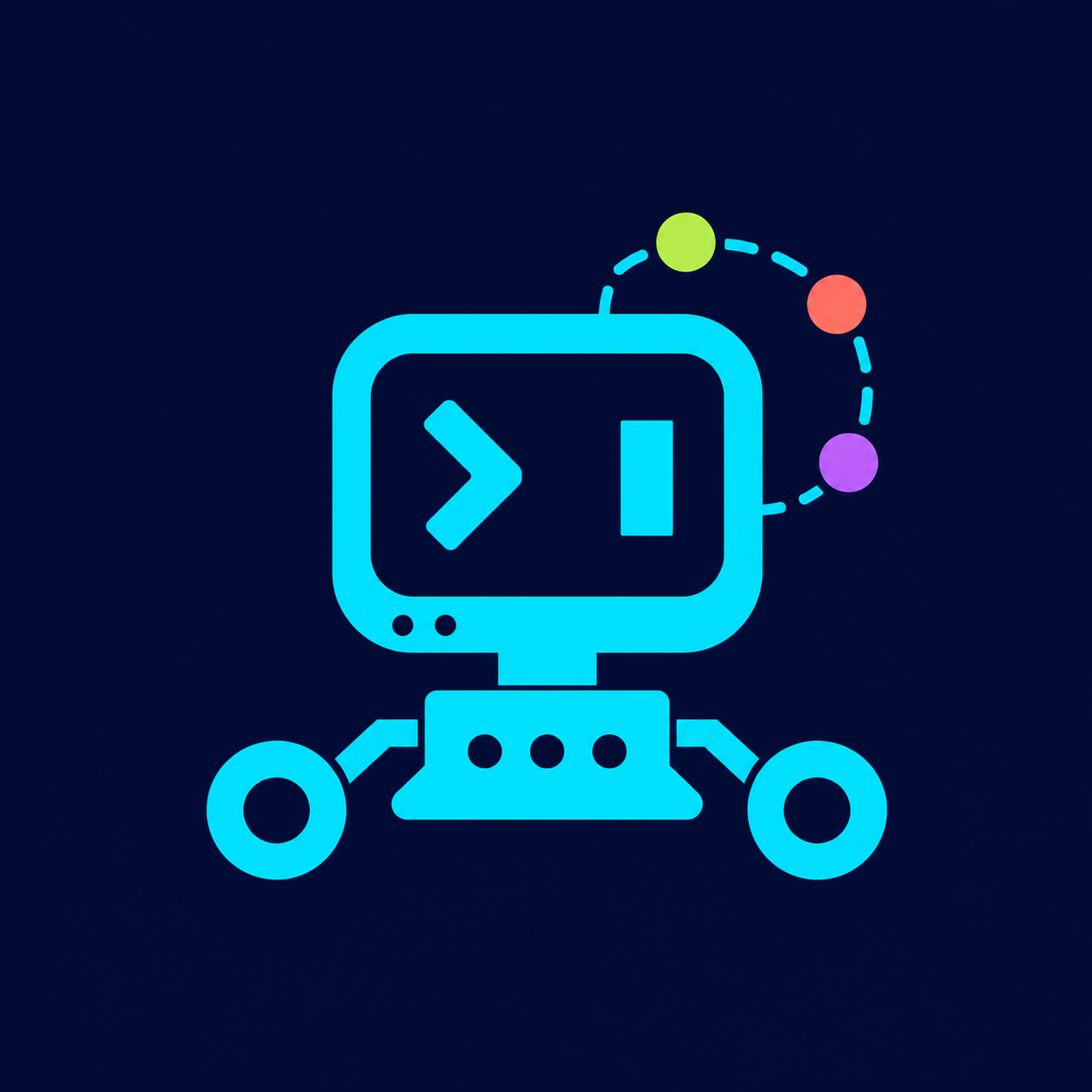}}}%
  \hspace{0.45em}%
  \href{https://github.com/tui-testing/tuicov}{%
    \raisebox{-0.30\height}{\includegraphics[height=0.50in]{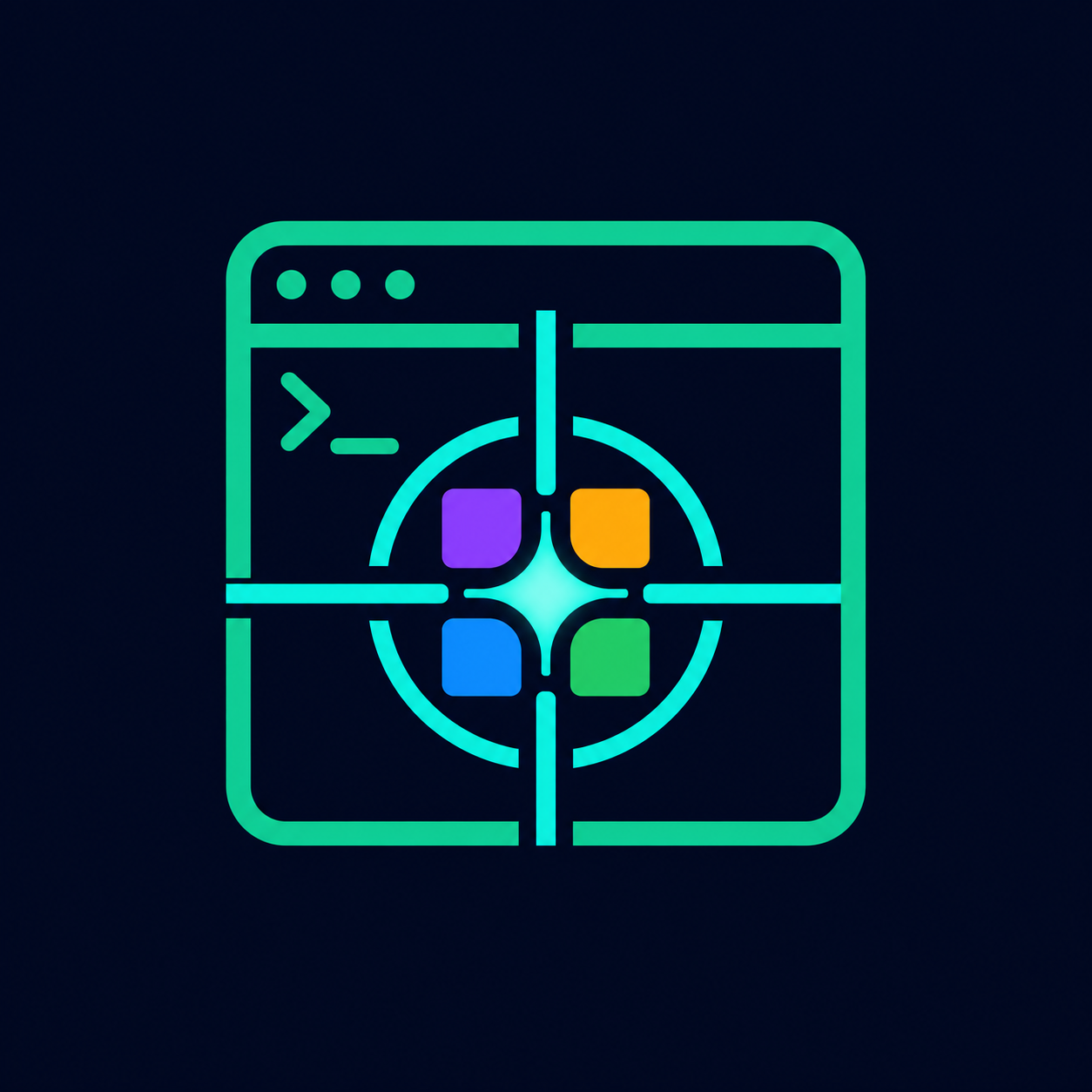}}}%
  \hspace{0.45em}Can LLMs Test Terminal User Interfaces?%
}{Can LLMs Test Terminal User Interfaces?}}
\author{%
Chao Peng\thanks{Chao Peng and Ruida Hu are co-first authors and contributed equally.} \\
University of Edinburgh \\
United Kingdom \\
{\small\texttt{chao.peng@acm.org}}
\And
Ruida Hu\textsuperscript{*} \\
Harbin Institute of Technology \\
Shenzhen, China \\
{\small\texttt{200111107@stu.hit.edu.cn}}
\And
Ajitha Rajan \\
University of Edinburgh \\
United Kingdom \\
{\small\texttt{arajan@exseed.ed.ac.uk}}
\AND
Tegawend\'e F. Bissyand\'e \\
University of Luxembourg \\
Luxembourg \\
{\small\texttt{tegawende.bissyande@uni.lu}}
\And
Jacques Klein \\
University of Luxembourg \\
Luxembourg \\
{\small\texttt{jacques.klein@uni.lu}}
\And
Cuiyun Gao \\
Harbin Institute of Technology \\
Shenzhen, China \\
{\small\texttt{gaocuiyun@hit.edu.cn}}
}

\maketitle

\begin{abstract}
\footnotesize
Terminal User Interfaces (TUIs) combine the stateful, screen-oriented behaviour of GUIs with terminal deployment and are now common in developer tools. Yet they lack a dedicated testing methodology. We survey 197 real-world TUI applications: only 12\% of test code exercises the interface, and 45\% of those tests never send input, checking a static frame instead. We turn these applications into a headless benchmark spanning ratatui/Rust, bubbletea/Go, textual/Python, and ink/TypeScript, packaging each as an instrumented Docker image. We record line and widget coverage where reliable, rendered terminal states, and crashes. Under equal wall-clock budgets, we compare four frontier LLMs with random exploration. No model dominates. Random is a strong time-budgeted baseline, but its crash advantage comes from higher throughput: per interaction, LLM guidance is more efficient and uniquely reaches input-gated faults. Automatically deriving launch inputs yields the largest practical gain, enabling applications that otherwise never start. Line coverage poorly predicts crash discovery, weakening it as a proxy for test effectiveness. Automated TUI testing is feasible but far from solved, and honest baselines matter more than model choice. We release the coverage tool \texttt{tuicov} at \url{https://github.com/tui-testing/tuicov} and the testing framework \texttt{tuibot} at \url{https://github.com/tui-testing/tuibot}.

\end{abstract}

\vspace{-1.2ex}
\begin{figure}[H]
\centering
\includegraphics[width=0.75\linewidth]{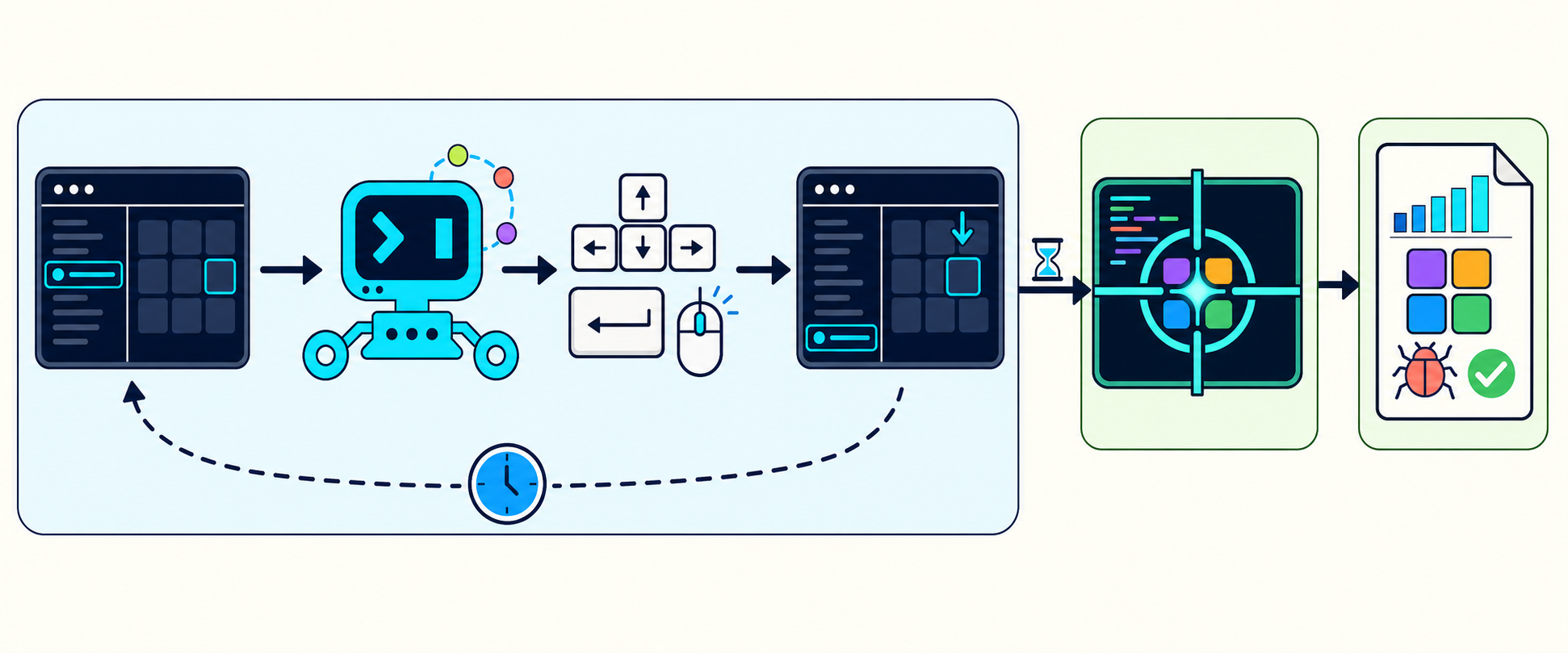}
\vspace{0.1em}

{\scriptsize
\textbf{Testing loop:} rendered TUI $\rightarrow$ \texttt{tuibot} $\rightarrow$ generate action $\rightarrow$ exercise TUI $\circlearrowleft$ until the budget is exhausted.\\[-0.1em] \textbf{Post-run:} \texttt{tuicov} measures execution $\rightarrow$ coverage and crash report.}

\captionsetup{font=scriptsize,skip=2pt}
\caption{Control flow of the released \texttt{tuibot} and \texttt{tuicov}
toolchain.}
\label{fig:tuibot-tuicov-overview}
\end{figure}

\section{Introduction}
\label{sec:intro}

Terminal User Interface (TUI) applications combine properties of command-line (CLI) and graphical (GUI) software without belonging to either. Like CLI tools they run inside a terminal emulator and are launched from a shell. Like GUI applications they present a stateful, screen-oriented interface with windows, panes, focus management, modal dialogs, and keyboard- and mouse-driven navigation. This class is large and growing fast. File managers, system monitors, database clients, git frontends, and, most recently, LLM coding agents are routinely shipped as TUIs, built atop frameworks such as ratatui~\cite{ratatui}, bubbletea~\cite{bubbletea}, textual~\cite{textual}, and ink~\cite{ink}.

\begin{figure}[t]
\centering
\includegraphics[width=0.8\textwidth]{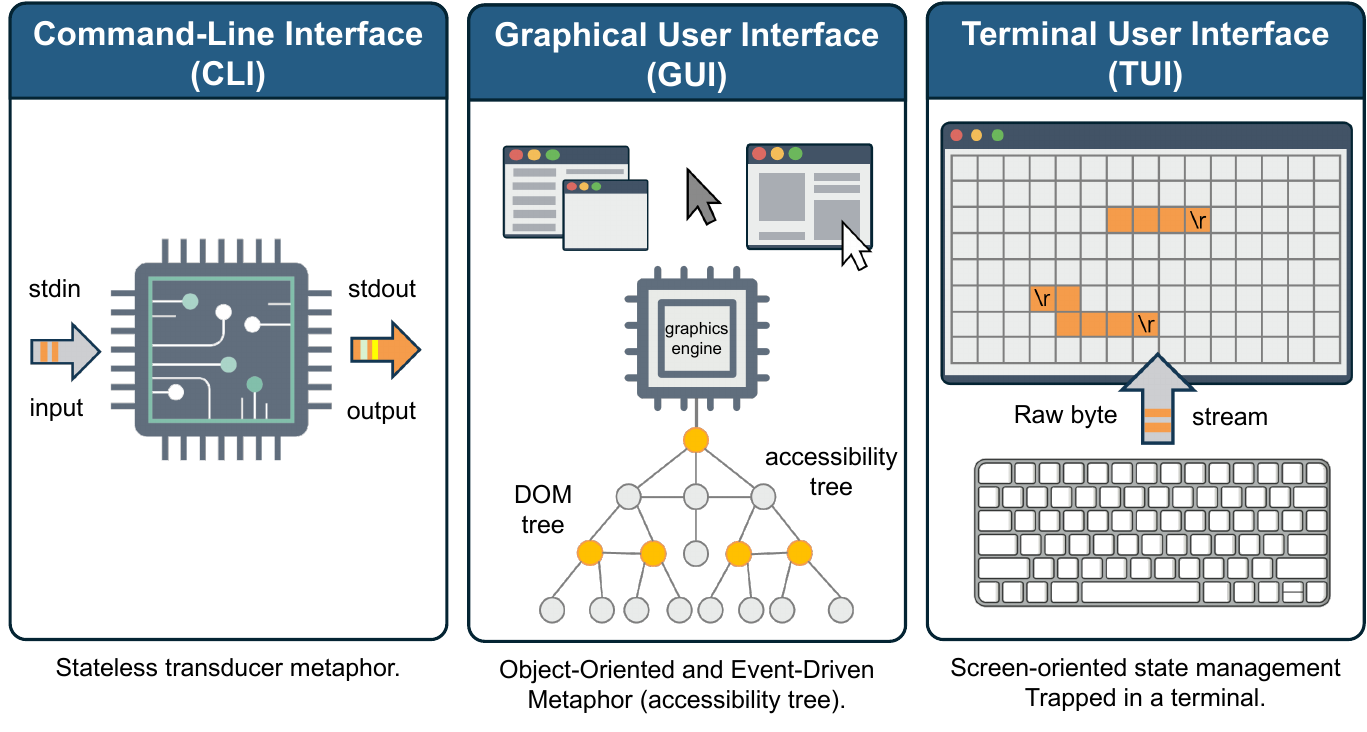}
\caption{A technical comparison of the three interface families. The CLI is a stateless transducer (\texttt{argv}/\texttt{stdin} in, \texttt{stdout} out); the GUI exposes state and events as addressable objects through an accessibility tree; the TUI exposes neither, presenting only a character grid rendered into a pseudo-terminal driven by a raw byte stream.}
\label{fig:interface-comparison}
\end{figure}

Despite their prevalence, TUI applications have no dedicated testing methodology, and crashes in TUI software remain both common and uninvestigated. Existing research targets either CLI programs, treated as pure input/output transducers whose correctness is judged by exit codes and stdout, or GUI programs, driven through accessibility trees and pixel-level widget hierarchies~\cite{su2017stoat,machiry2013dynodroid,mao2016sapienz}. Neither model transfers to a TUI. CLI techniques discard all stateful, screen-oriented behaviour, which is the very layer where TUI bugs manifest. GUI techniques assume rich accessibility metadata, namely widget identifiers, roles, and bounding boxes, that a character-grid terminal does not expose. There is no DOM, no accessibility tree, and no coordinate system beyond a two-dimensional array of cells rendered into a pseudo-terminal. A TUI's entire observable state is a character grid, and its behaviour is driven by raw key, mouse, and resize events. A comparison of the three interfaces is shown in Figure~\ref{fig:interface-comparison}. No existing fuzzing, exploration, or test-generation technique was designed for that interaction model. The consequence is a practical gap. TUI applications crash, users file bug reports, and no automated technique exists to find, reproduce, or prevent those failures.

Two questions motivate this paper. First, \emph{how well are TUIs tested today?} We surveyed the test suites of 197 real-world TUI applications and found the interface layer doubly under-tested: only about 12\% of all test code exercises the terminal/widget layer at all, and of the tests that do, nearly half never send a single input event (Section~\ref{sec:motivation}). Second, \emph{do the LLM-driven exploration and test-generation techniques now reshaping CLI and GUI testing~\cite{deng2023titanfuzz,schafer2024llmunit,liu2024guiGPT} transfer to TUIs?} Our results suggest they do not transfer cleanly. The relationship between code coverage and crash detection in TUI applications diverges from the pattern observed in CLI and GUI domains, where higher coverage tends to predict more faults found~\cite{strecker2008relationships,yuan2011gui,silva2023flacoco}. In TUIs, coverage and fault-finding decouple. That finding undermines the standard practice of using line coverage as a proxy for test effectiveness and motivates a TUI-specific coverage metric that aligns with crash discovery. We focus on \emph{crash bugs}. Silent logic errors and performance regressions matter, but they require ground truth that does not yet exist for TUIs, whereas crashes are unambiguous, automatically detectable, and, as our bug corpus shows, a substantial and underexplored failure mode in practice.

We make the following contributions:

\begin{itemize}
\item \textbf{An open, multi-language TUI benchmark.} 197 real-world applications spanning the four dominant frameworks and languages (ratatui/Rust, bubbletea/Go, textual/Python, ink/TypeScript), each packaged as a per-instance, instrumented Docker image that builds and runs headlessly (Section~\ref{sec:benchmark}).
\item \textbf{A kill-resilient instrumentation toolkit.} \texttt{tuicov}\footnote{\url{https://github.com/tui-testing/tuicov}} is a language-agnostic coverage tool that records both native line coverage and an experimental widget-coverage signal from an interactive application that must be terminated rather than allowed to exit (Section~\ref{sec:benchmark}).
\item \textbf{An empirical study and open-source test framework.} \texttt{tuibot}\footnote{\url{https://github.com/tui-testing/tuibot}} implements the frontier-LLM and exploration strategies evaluated on this benchmark, answering three research questions about model capability, technique comparison, and the coverage-fault relationship  (Sections~\ref{sec:rqs}-\ref{sec:results}).
\end{itemize}

Our findings counsel caution about the anticipated benefits of LLM-based TUI testing. No single model achieves consistent superiority across applications. A model-free random baseline remains competitive throughout. The principal determinant of practical effectiveness proves to be the correct derivation of launch inputs, without which input-dependent applications cannot be exercised at all. Most consequentially, conventional line coverage is a poor predictor of crash discovery in TUI software, which reinforces the case for a domain-specific coverage criterion. We release \texttt{tuicov} and \texttt{tuibot} as open-source tools to support TUI testing as an independent research area.

\section{Motivation}
\label{sec:motivation}

\textbf{Why TUIs need their own testing approach.}

To understand why existing testing techniques fall short for TUIs, consider a concrete behaviour from a file-manager TUI: pressing \texttt{j} moves the cursor down, and pressing \texttt{Enter} opens the highlighted entry. This single, familiar interaction exposes a fundamental mismatch with both CLI and GUI testing models, as illustrated in Figure~\ref{fig:interface-comparison}.

\emph{In the CLI model}, the interaction cannot be expressed at all. CLI testing treats a program as a pure input/output transducer: a test fixes \texttt{argv} and \texttt{stdin}, captures \texttt{stdout}/\texttt{stderr} and an exit code, and compares the result against an expected value. A sequence such as ``move cursor down, then open entry'' has no representation in this model. It unfolds across time against a persistent, on-screen state that a single-invocation transducer cannot capture.

\emph{In the GUI model}, the same behaviour is straightforwardly testable, because a graphical application exposes an accessibility tree of named, queryable objects. A test driver locates the file list by its accessibility role, reads its \texttt{selectedRow} property, synthesises a keypress or mouse click to advance the selection, and then queries \texttt{row.isSelected} or awaits an \texttt{activated} event. State and events are first-class, named entities that the harness can inspect and subscribe to directly.

\emph{In the TUI model}, neither of these handles exists. The entire observable state of a TUI is a two-dimensional grid of characters rendered into a pseudo-terminal; input is the raw byte stream \texttt{"j\textbackslash r"}. There is no queryable property for ``which row is highlighted'', that information is encoded implicitly as the single line drawn with an inverted SGR (Select Graphic Rendition) attribute somewhere within an $H\times W$ grid of cells. To verify that the cursor moved, a test must write the bytes into the pseudo-terminal (PTY), wait for the screen to repaint, scan the full cell grid for the styled run, and decode the file path from its glyphs. To verify that an entry was opened, it must diff the entire frame against a prior state. There is no widget to query and no event to await, with only bytes in and a repainted screen out. This is precisely the layer that existing CLI- and GUI-oriented techniques bypass, and the layer that any automated TUI tester must be capable of driving directly.

\begin{figure}[p]
\centering
\begin{subfigure}[b]{0.3\textwidth}
\centering
\includegraphics[width=\textwidth]{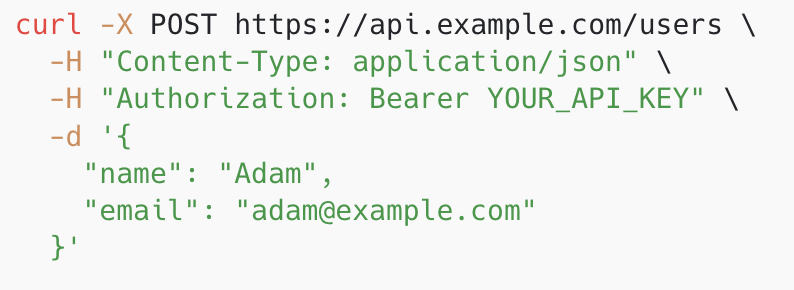}
\caption{CLI (\texttt{curl}).}
\label{fig:same-task-cli}
\end{subfigure}

\vspace{0.8em}

\begin{subfigure}[b]{\linewidth}
\centering
\includegraphics[width=\linewidth]{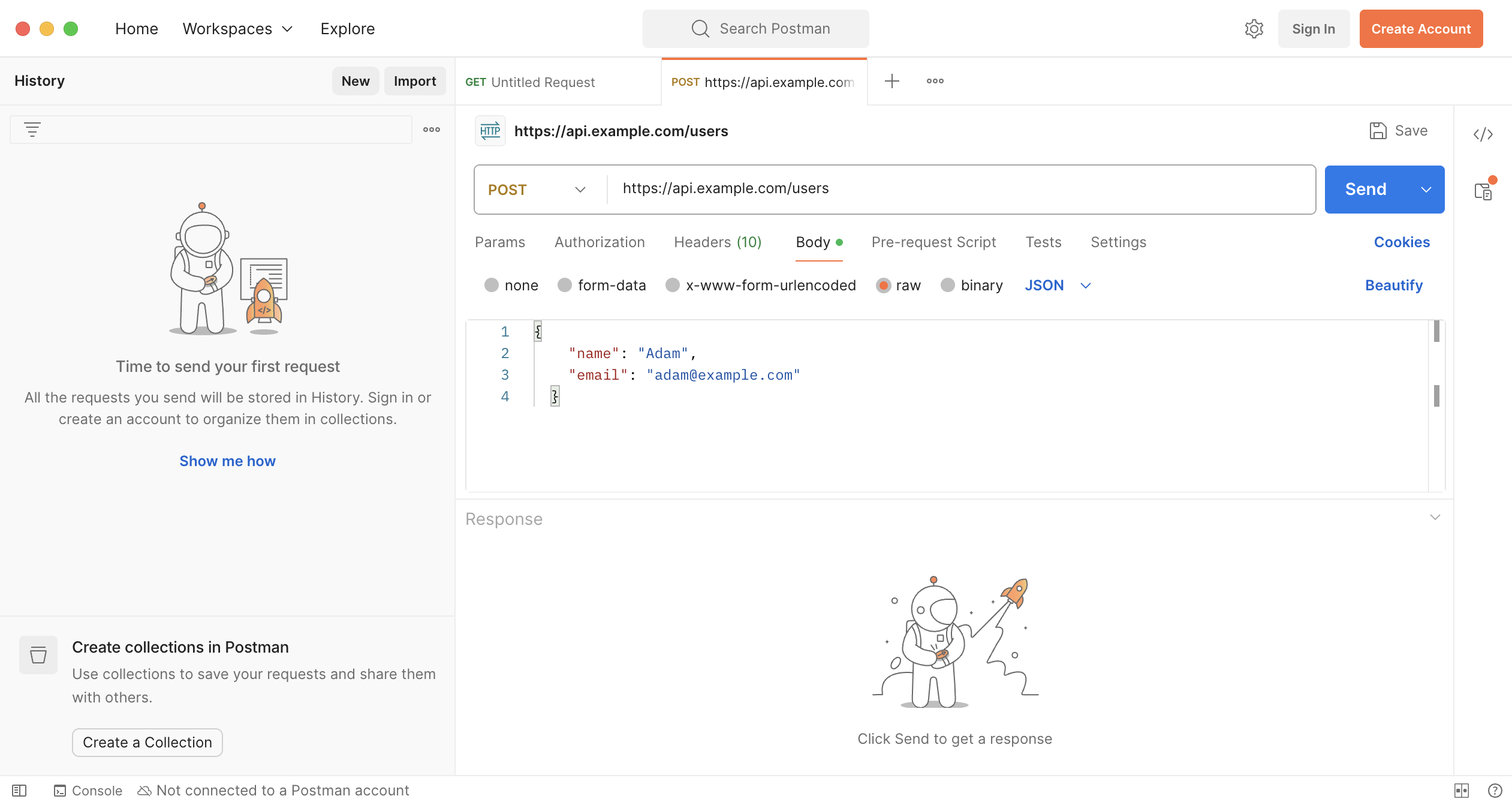}
\caption{GUI (Postman).}
\label{fig:same-task-gui}
\end{subfigure}

\vspace{0.8em}

\begin{subfigure}[b]{\linewidth}
\centering
\includegraphics[width=\linewidth]{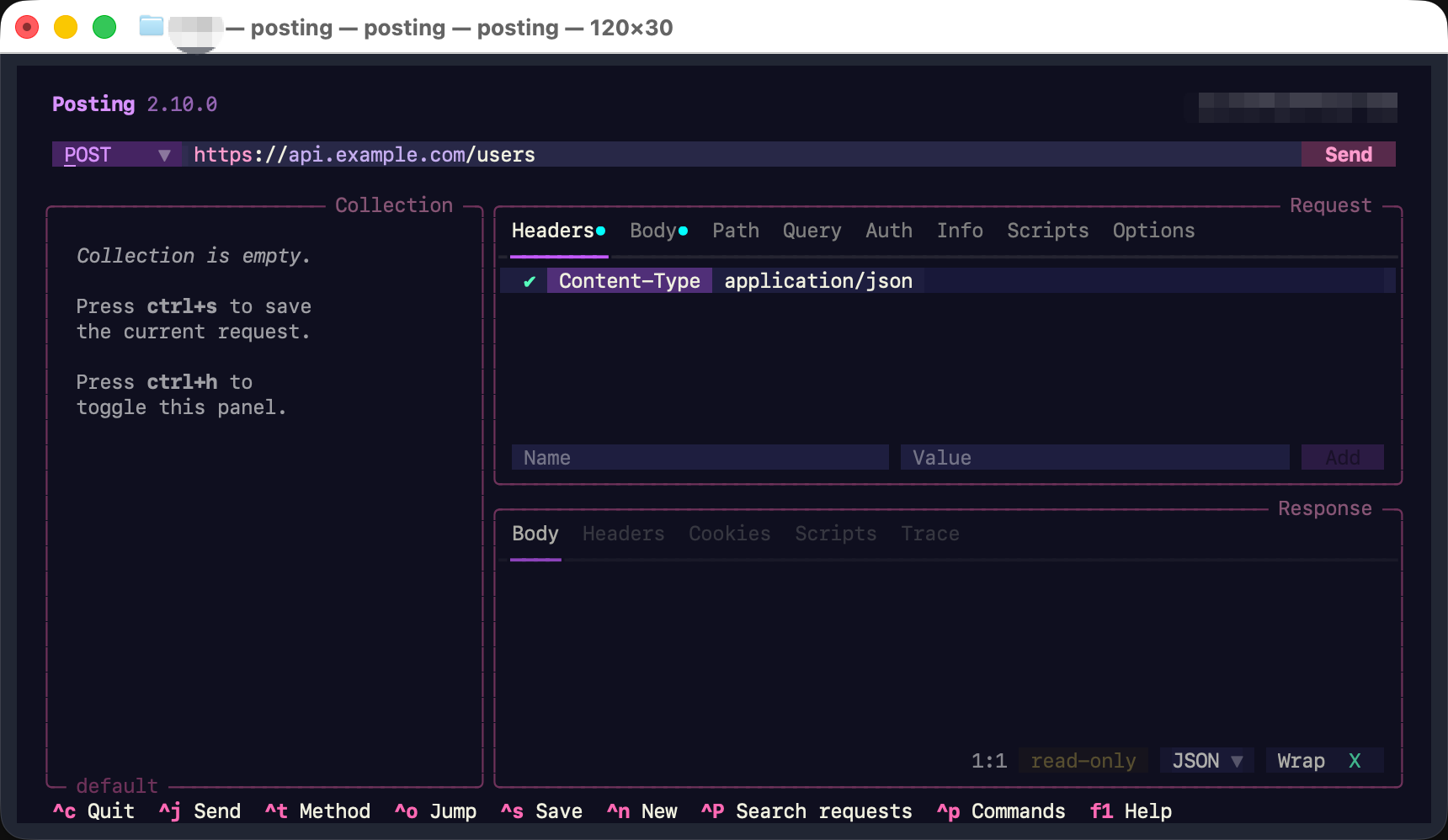}
\caption{TUI (Posting).}
\label{fig:same-task-tui}
\end{subfigure}
\caption{The same task (issuing an authenticated \texttt{POST} to
\texttt{/users} with a JSON body) across the three interface families.
The CLI (\subref{fig:same-task-cli}) encodes the whole request as one
\texttt{argv}/\texttt{stdin} invocation; the GUI
(\subref{fig:same-task-gui}) exposes named fields and buttons as
addressable objects; the TUI (\subref{fig:same-task-tui}) renders the
same fields as a character grid in a terminal, driven entirely by
keystrokes.}
\label{fig:same-task}
\end{figure}

Figure~\ref{fig:same-task} reinforces this point with a richer example: issuing an authenticated \texttt{POST} of a JSON user record, realised across all three interface families.

\textbf{TUIs are poorly tested in practice.}
TUI application frameworks provide testing harness and developers can write TUI-level tests with it. Figure~\ref{fig:twig-test} shows a representative example from \texttt{twig}, a JSON-explorer TUI in our benchmark, written against textual's \texttt{Pilot} driver. The harness boots the app inside a headless pseudo-terminal (\texttt{run\_test}), synthesises keystrokes (\texttt{pilot.press}), and, crucially, lets the test query live widget objects (\texttt{query\_one}, \texttt{screen.focused}, \texttt{OptionList.highlighted}) instead of decoding glyphs from the screen buffer. This is exactly the ``move cursor, then assert which row is highlighted'' interaction from our running example, made expressible only because the framework re-exposes the named state and events that the terminal itself discards. Every TUI ecosystem ships some version of this affordance (ratatui's \texttt{TestBackend}, bubbletea's \texttt{teatest}, ink's \texttt{ink-testing-library}); the question is how much developers actually use it.

\begin{figure}[htbp]
\begin{lstlisting}[language=Python]
async with TwigApp(SAMPLE).run_test() as pilot:
    nav = app.query_one("ColumnNavigator")
    # type a path into the search bar and jump to it
    await pilot.press("/")
    await pilot.press(*'.regions["us-east-1"].vpcs[0]')
    await pilot.press("enter")
    # assert on widget state, not on screen glyphs
    assert "col-" in str(app.screen.focused.id)
    # search "available", then step matches with n / N
    await pilot.press("/"); await pilot.press(*"available")
    await pilot.press("enter")
    first = nav_selected(nav)
    await pilot.press("n")          # next match
    assert nav_selected(nav) != first
    await pilot.press("N")          # previous match
    assert nav_selected(nav) == first
\end{lstlisting}
\caption{A TUI-level test from \texttt{twig} using textual's \texttt{Pilot} harness (condensed from \texttt{tests/test\_integration.py}). The harness drives the app through keystrokes and asserts on queryable widget state, which is the affordance that makes ``press a key, check which row is highlighted" testable.}
\label{fig:twig-test}
\end{figure}

To establish that this gap has real-world consequences, we surveyed the test suites of all 197 benchmark applications, classifying every test file (excluding vendored and dependency code) \sloppy{as either \emph{TUI-level} (exercising the terminal or widget layer via the framework's TUI test harness) or \emph{non-TUI}, in the spirit of prior test-suite effectiveness studies~\cite{inozemtseva2014coverage}. Classification was LLM-assisted and verified by manual inspection; all counts are computed deterministically.}

\textbf{Most applications barely test their interface.} Of the 197 applications surveyed, only 76 (38.6\%) include \emph{any} TUI-level tests; 35.0\% have tests that never touch the UI layer at all; and 26.4\% ship no tests whatsoever. As shown in Table~\ref{tab:survey-share}, only approximately 12\% of all test code exercises the TUI itself. The remainder targets parsers, configuration, data layers, and business logic. Even textual, the ecosystem with the most mature TUI testing infrastructure, allocates only a quarter of its test files to the UI. The \emph{Median app share} column computes each application's own TUI-test fraction and takes the median across applications, describing the \emph{typical} app rather than the aggregate. The three can diverge sharply, which is precisely why we report all of them: ink, for instance, shows 18.3\% of files but a median of 0\%, because its pooled figure is inflated by a single exhaustively tested outlier (\texttt{gemini-cli}) while the median ink application has no TUI test at all.

\begin{table}[htbp]
\caption{Fraction of test effort targeting the TUI layer, by framework.}
\label{tab:survey-share}
\centering
\begin{tabular}{lrrr}
\hline
\textbf{Framework} & \textbf{Apps w/ TUI test} & \textbf{TUI files / all} & \textbf{Median app share} \\
\hline
ratatui   & 32\% & 8.9\%  & 7.5\% \\
bubbletea & 43\% & 10.0\% & 6.2\% \\
textual   & 41\% & 25.4\% & 15.6\% \\
ink       & 43\% & 18.3\% & 0.0\% \\
\hline
\textbf{All} & \textbf{38.6\%} & \textbf{12.3\%} & --- \\
\hline
\end{tabular}
\end{table}

\textbf{The same gap at the test-case level.} Table~\ref{tab:survey-cases} re-counts at the granularity of individual test cases (test functions). The 785 TUI-level test files resolve to 8,353 distinct TUI-level test cases (10.6 test cases per file in average) , of which only 5,213 (62\%) are \emph{interactive}.

\begin{table}[htbp]
\caption{TUI-level testing at the granularity of individual test cases (test functions), by framework. \emph{Interactive} cases belong to a test that sends at least one key, mouse, or resize event.}
\label{tab:survey-cases}
\centering
\begin{tabular}{lrrr}
\hline
\textbf{Framework} & \textbf{TUI test cases} & \textbf{Interactive} & \textbf{Interactive \%} \\
\hline
ratatui   & 1{,}376 & 511   & 37\% \\
bubbletea & 3{,}820 & 2{,}841 & 74\% \\
textual   & 1{,}583 & 1{,}357 & 86\% \\
ink       & 1{,}574 & 504   & 32\% \\
\hline
\textbf{All} & \textbf{8{,}353} & \textbf{5{,}213} & \textbf{62\%} \\
\hline
\end{tabular}
\end{table}

\textbf{The TUI tests that do exist are shallow.} Among the 785 TUI-level test files found across the 76 applications that include them, only 55\% (430 test files) are genuinely \emph{interactive}, that is, they feed key, mouse, or resize events and assert on the resulting state. A further 24\% are \emph{render-only} tests that draw a single static frame and diff the buffer, and the remaining 21\% are \emph{snapshot-only} tests that compare against a golden frame without sending any input. In total, 45\% of TUI test files never send a single input event. Sixteen of the 76 applications (21\%), including widely used tools such as \texttt{bottom}, \texttt{bandwhich}, \texttt{kmon}, and \texttt{television}, have TUI test suites that are composed \emph{entirely} of static renders or snapshots. The interactive tests are also thin. The median interactive file uses only nine input actions, and 11\% use just one or two. A file with one or two input actions typically presses a single key and checks a single outcome, exercising none of the multi-step navigation, focus transitions, modal dialogs, or edge-case key sequences that characterize real use.

The TUI interface layer is therefore doubly under-tested: it receives a small fraction of overall test effort, and the majority of that effort is static. This is the gap that an automated TUI exploration approach is positioned to fill.

\section{Benchmark and Infrastructure}
\label{sec:benchmark}

\begin{figure}[t]
\centering
\includegraphics[width=0.95\textwidth]{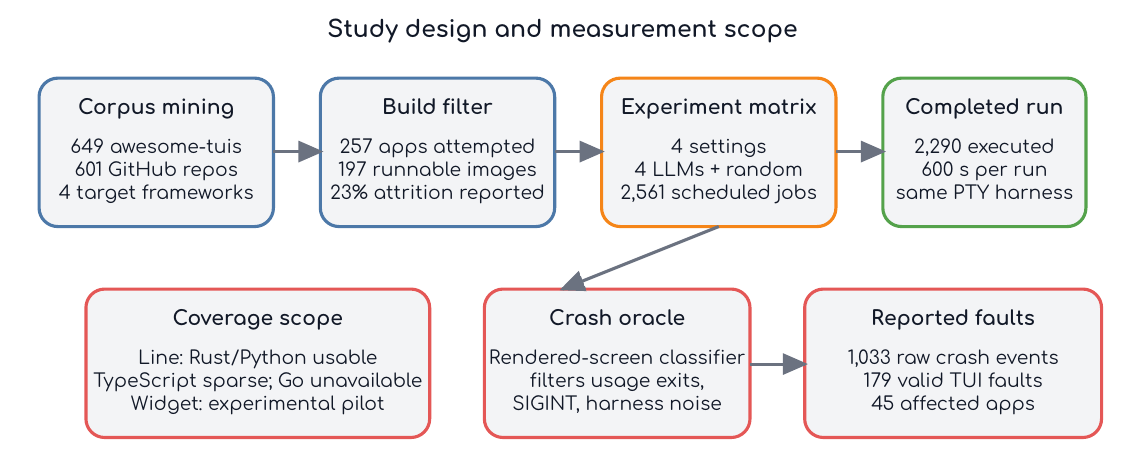}
\caption{Overview of the completed study design and the measurement scope used in the paper. The benchmark starts from the mined TUI corpus, filters to 197 headlessly runnable applications, runs the four exploration settings under the same PTY harness, and reports only measurements that are valid in the completed run.}
\label{fig:study-design}
\end{figure}

Figure~\ref{fig:study-design} gives a roadmap of the benchmark and the measurement scope used throughout the paper. The top row traces the construction and execution pipeline: we mine a corpus of real TUIs and resolve their frameworks, filter to the applications that build and run headless as per-instance instrumented images, schedule the four-setting experiment matrix across the model panel, and execute it. The bottom row records what the completed run can measure: the cross-language coverage scope, the rendered-screen crash oracle, and the resulting fault counts.

\subsection{Frameworks and Applications}

To ground our choice of frameworks in evidence rather than intuition, we first mined every TUI listed in \texttt{awesome-tuis}\footnote{\url{https://github.com/rothgar/awesome-tuis}}, an actively maintained collection of TUI applications with approximately 20,000 GitHub stars. In this collection, there are 649 projects across 13 categories. We fetched each repository's evidence files (README, \texttt{pyproject.toml}, \texttt{Cargo.toml}, \texttt{package.json}, \texttt{go.mod}, \dots), and asked three LLMs (Claude-Sonnet-4.5, Claude-Opus-4.7, GPT-5.4) to identify each project's underlying TUI library from that evidence. The three models agreed on the same library for 76\% of resolvable projects and never once named three different libraries, giving us confidence in the resulting tally. The result is a near-monolithic leader \emph{within each language}: of projects whose library could be identified, Rust overwhelmingly uses \textbf{ratatui} (78\%), Go uses \textbf{bubbletea} (the clear plurality), Python uses \textbf{textual}, and TypeScript uses \textbf{ink}. We therefore target exactly these four frameworks, one per mainstream language, as the dominant choice each ecosystem has converged on. From these ecosystems we assemble a suite of real-world applications spanning a wide popularity range, from 100k-star coding agents (codex, gemini-cli) down to niche hobby tools.

\subsection{Per-Instance Instrumented Images}

As there is no existing unified tool to measure how thoroughly a TUI's interface is exercised across all four target languages and frameworks, we implement \textbf{\texttt{tuicov}}\footnote{\url{https://github.com/tui-testing/tuicov}}, an open-source, language-agnostic coverage toolkit that underpins the benchmark and the measurements throughout this paper. Given an application's source, \texttt{tuicov} performs a four-stage \emph{instrument $\rightarrow$ build $\rightarrow$ run $\rightarrow$ report} pipeline. It first parses the source and statically detects \emph{widget sites}, the source locations where the framework's UI components (e.g., ratatui, bubbletea, textual, and ink widgets) are constructed or rendered, and injects a lightweight runtime probe at each one. The instrumented app is then built under each language's native line-coverage facility (\texttt{cargo llvm-cov} for Rust, \texttt{go tool covdata} for Go, coverage.py for Python, and \texttt{c8} for TypeScript). At runtime the probe appends one record per widget hit to a log, while the native facility records executed lines. After a driven session, a language-agnostic reporter merges the two raw sources into a single \emph{unified report} that pairs native \textbf{line coverage} with TUI-specific \textbf{widget coverage} per file. This dual signal     (ordinary code coverage alongside an interface-aware one) is what lets us reason about \emph{interface} exploration rather than mere code execution.

The benchmark is realized as per-instance, \texttt{tuicov}-instrumented Docker images that run in a plain headless Linux environment. A single Dockerfile family (one per language) clones the app at the target commit, applies instrumentation in place, builds under native coverage when the runtime supports it, and ships a runnable image.

Of 257 apps attempted at the most-recent commit, \textbf{197} produced a runnable image. A subset of arbitrary apps do not build or run headless, whether from missing system dependencies, display/daemon requirements, or compile failures.

\subsection{Execution Harness and Coverage Instrumentation}

Each run is driven by \texttt{tuibot}\footnote{\url{https://github.com/tui-testing/tuibot}}, our open-source automated TUI testing framework, through a pseudo-terminal at a fixed terminal size. The harness waits for the rendered grid to stabilize after each action, records the character-grid state, and sends key, mouse, and resize events as raw terminal input. The same harness backs all four exploration settings we study: (1) \emph{random} (model-free input as a baseline), (2) \emph{llm-guided} (an LLM picks each action at runtime), (3) \emph{llm-guided-derived} (llm-guided seeded with automatically derived launch inputs), and (4) \emph{llm-generated} (an LLM reads the source and emits test scenarios that steer the explorer), defined in detail in Section~\ref{sec:rqs}.

A key subtlety distinguishes TUI testing from ordinary unit testing. The exploration harness \emph{terminates} each session when the time budget is used up, but most native coverage tools only flush on a clean exit. This makes line coverage unavailable for Go as it does not support extracting line coverage when the app is terminated by external signals.

\subsection{A Content-Aware Crash Oracle}
\label{sec:oracle}

A consequential lesson from driving 197 apps is that \textbf{``process exited non-zero'' is not a usable crash oracle for TUIs}. Across the capability run, 1{,}033 crash events were recorded, but only 179 (17.3\%) are valid TUI-level faults after our examination. The remaining 82\% are noise of a few distinct kinds (Section~\ref{sec:results}). These include an app printing a usage message and exiting because it needed required arguments, a graceful response to Ctrl+C, a \texttt{docker run} name clash, a PTY ioctl failure, and, as the single largest category, a coverage-runtime emit failure that turns a clean exit into a non-zero one. An exit-code oracle counts all of these as crashes.

We therefore classify each crash event on its \emph{rendered terminal screen} rather than its exit code. A content-aware classifier scans the post-termination grid for a rendered traceback, an on-screen runtime exception, a Rust/Go panic, or a fatal signal (SIGSEGV/SIGABRT), and hard-excludes harness artifacts (coverage-emit failures, docker races, PTY errors, SIGKILL-at-budget, usage/\texttt{--help} exits). Events with no decisive on-screen signal are adjudicated individually: two of the authors independently labeled all 233 ambiguous crash events as valid TUI-level faults, harness or environment noise, or undecidable. The annotators reached 93\% raw agreement, and the remaining disagreements were resolved through discussion until consensus. Per-app crash \emph{saturation}, the fraction of an app's runs that crash, together with the median number of steps to the crash, further separates input-invariant startup and environment failures (high saturation, crash at step~1) from interaction-reached faults (mid saturation, many steps). A direct implication for future TUI fault studies is that we adopt a rendered-screen oracle rather than counting raw exits.

\section{Research Questions and Study Design}
\label{sec:rqs}

We investigate three research questions. Across all RQs we evaluate on TUI applications spanning multiple languages and frameworks (Rust/ratatui, Go/bubbletea, Python/textual, TypeScript/ink) to assess generalization, and all techniques drive each app through a pseudo-terminal, in the manner of end-to-end terminal testing tools~\cite{tuitest}, running through the same instrumented-container harness so their coverage and crash measurements are directly comparable.

\paragraph{RQ1: How do frontier LLMs perform on testing TUI applications?}
We compare four frontier LLMs (\texttt{Claude-Opus-4.8}, \texttt{GPT-5.5}, \texttt{Gemini-3.5-Flash}, and \texttt{DeepSeek-V4-Pro}) using a fixed LLM-guided exploration strategy. We measure code coverage achieved, the crashes uncovered, and the \textbf{token cost} each model incurs, with budgets equalized in \textbf{wall-clock time per session} so every model drives each subject app for the same exploration window regardless of its latency or pricing. Because the budget is fixed in time rather than in steps, a model's per-step throughput, namely how many actions it can issue before the clock runs out, is itself part of what RQ1 measures.

\paragraph{RQ2: How do random exploration, LLM-guided exploration, and LLM-based test generation compare?} We compare four settings: (1) random exploration as a model-free baseline, (2) LLM-guided runtime exploration, (3) LLM-guided exploration augmented with automatically derived launch inputs, and (4) LLM-based test generation from source code. We measure coverage and crashes, again with budgets equalized in wall-clock time per session, and, for the crash comparison, report effectiveness \emph{normalized by interaction} (valid crashes per unit of input) rather than by raw count, because a time-equalized budget gives each setting a very different number of steps.

\paragraph{RQ3: Does code coverage correlate with fault-finding in TUI applications?} We analyze the relationship between code coverage and the number of \emph{valid} faults discovered across sessions and applications, using the adjudicated crash set (Section~\ref{sec:results}) as observed fault evidence and accounting for the session-truncation confound that a crash introduces by ending an episode.

\subsection{Techniques (Settings)}

The comparison axis comprises four settings.
\begin{itemize}
  \item []\textbf{(1) random:} Model-free random input, which serves as the baseline and is analogous to random/monkey testing in the GUI domain~\cite{machiry2013dynodroid,su2017stoat,choudhary2015monkey}.
  \item[]\textbf{(2) LLM-guided:} An LLM selects each action at runtime taking the current terminal screen (characters shown in terminal) as input.
  \item[]\textbf{(3) LLM-guided-derived:} The LLM-guided setting augmented with an LLM that derives CLI arguments and input fixtures so that input-dependent applications launch.
  \item[]\textbf{(4) LLM-generated:} An LLM reads source code and emits natural-language test scenarios that subsequently steer the LLM-guided explorer.
\end{itemize}

Each LLM setting runs across the model panel; random is model-independent. To account for the stochasticity of both random and LLM-guided exploration, every configuration is repeated \textbf{three times} per app, and all reported metrics are averaged over the three runs. Each run receives a \textbf{600\,s} wall-clock budget, equalized across settings. Exploration relaunches the app on exit/crash until the budget is spent, so coverage and crashes accumulate across episodes. To isolate each arm, the LLM settings employ \textbf{no random fallback}. A failed LLM call is retried and then fails the run rather than degrading silently to random, so that every step attributed to an LLM arm constitutes a genuine model decision.

\subsection{Metrics}

We record three signals per run. \textbf{Line coverage} and \textbf{widget coverage} are harvested from the unified report after each session (Section~\ref{sec:benchmark}), but the results scope each signal to the runtimes where the completed run produced valid data. For crashes we do \emph{not} report raw exit-non-zero counts, because, as Section~\ref{sec:results} shows, those are dominated by environment and harness noise. We instead classify every crash event on its post-termination terminal screen into \emph{valid} TUI-level faults and noise, and we report valid crashes both per run and \textbf{per unit of interaction} (per~1{,}000 steps), together with the count of unique \texttt{(app, fault)} signatures. For every LLM run we additionally record \textbf{token usage} (input tokens, output tokens, and number of model calls), so that effectiveness can be weighed against cost.

\section{Results}
\label{sec:results}

Our study comprises a unified run of 197 applications under four settings and four models, with every configuration repeated three times per app and all reported numbers averaged over the three runs. Before answering the RQs, we scope the measurements. Crash results use the \emph{adjudicated} valid-fault set rather than raw exit codes. Line coverage is reported for Rust, Python, and TypeScript, not for Go due to framework limitations as discussed in Section~\ref{sec:benchmark}.

\subsection{RQ1: No model dominates, and capability is decoupled from cost}

Across the model panel, no single LLM leads on every metric. Within each setting$\times$language cell, the best model on line coverage is rarely the best on crashes, and the four models are close on unique-fault discovery (Gemini~27, Claude~24, GPT-5.5~18, and DeepSeek~16 unique valid faults). The same non-dominance holds on coverage: broken down by model (Table~\ref{tab:results-setting-model}), \texttt{Gemini-3.5-Flash} and \texttt{GPT-5.5} lead line coverage in every setting while the costliest model, \texttt{Claude-Opus-4.8}, tops none. What separates the models sharply is \textbf{token cost}, not capability (Table~\ref{tab:tokens-model}). Under the same 600\,s budget, the models differ by an order of magnitude in tokens consumed and in how many steps they manage to issue. \texttt{Claude-Opus-4.8} spends $\sim$150k tokens/run over $\sim$53 model calls, while \texttt{GPT-5.5} reaches a comparable outcome with $\sim$40k tokens, and \texttt{DeepSeek-V4-Pro} with only $\sim$24k. Because the budget is fixed in \emph{time}, slow or verbose models take fewer steps before the clock expires, which, as RQ3 shows, is the dominant driver of measured effectiveness.

\begin{table}[htbp]
\caption{Per-model cost under the equalized 600\,s budget (LLM settings, weighted by runs)}
\label{tab:tokens-model}
\centering
\begin{tabular}{lrrrr}
\hline
\textbf{Model} & \textbf{Tokens/run} & \textbf{Calls/run} & \textbf{Tok/call} \\
\hline
Claude-Opus-4.8        & 150{,}289 & 52.7 & 2{,}853 \\
Gemini-3.5-Flash       & 98{,}146  & 51.2 & 1{,}916 \\
GPT-5.5                & 39{,}777  & 17.5 & 2{,}274 \\
DeepSeek-V4-Pro        & 23{,}505  & 13.0 & 1{,}813 \\
\hline
\end{tabular}
\end{table}

\subsection{RQ2: Coverage and crash-finding by setting}

\paragraph{Coverage.} Table~\ref{tab:results-setting} summarizes per-app union line coverage (the union of lines covered across an app's runs) on the usable line-coverage subset, together with per-run token cost and steps. Within that scope, \texttt{LLM-guided-derived} leads on line coverage (30.4\% vs.\ 26--28\% for the other arms). The gain is concentrated in getting input-hungry apps to start at all. Plain arms reach 0\% on these apps that exit immediately without arguments or fixtures, and input derivation rescues them. Smarter per-step reasoning alone (LLM-guided vs.\ random) yields little additional coverage. Random's 28.4\% is competitive with both non-derived LLM arms despite issuing only model-free input. The determining factor is input derivation, not action selection.

\begin{table}[htbp]
\caption{Coverage and cost by setting. Line~\% is per-app union coverage on the usable line-coverage subset; tokens are per LLM run.}
\label{tab:results-setting}
\centering
\begin{tabular}{lrrr}
\hline
\textbf{Setting} & \textbf{Line \%} & \textbf{Tokens/run} & \textbf{Steps} \\
\hline
random                       & 28.4 & --- & 784 \\
LLM-guided                   & 26.4 & 73{,}005 & 35 \\
\textbf{LLM-guided-derived}  & \textbf{30.4} & 75{,}988 & 36 \\
LLM-generated                & 26.6 & 94{,}371 & 30 \\
\hline
\end{tabular}
\end{table}

\paragraph{Coverage by model.} Table~\ref{tab:results-setting-model} breaks the same per-app union coverage down by model within each LLM setting, alongside the per-run token cost. Two patterns hold across all three settings. First, \textbf{capability does not track cost}: \texttt{Gemini-3.5-Flash} and \texttt{GPT-5.5} reach the highest line coverage, yet \texttt{GPT-5.5} does so at roughly a quarter of \texttt{Claude-Opus-4.8}'s token spend and \texttt{Claude} never leads a single cell despite being the most expensive. Second, \textbf{input derivation helps every model}: each of the four gains 3--8 points moving from \texttt{LLM-guided} to \texttt{LLM-guided-derived}, confirming that the derived-input effect is a property of the setting rather than of one strong model.

\paragraph{Widget coverage.} Alongside line coverage we collected the interface-level \emph{widget-coverage}. There, widget coverage is \emph{flat across settings}: 33.3\% (random), 34.2\% (\texttt{LLM-guided}), 37.1\% (\texttt{LLM-guided-derived}), and 35.1\% (\texttt{LLM-generated}). The ordering echoes line coverage, with input derivation improving coverage up a few points and action-selection strategy otherwise making little difference.

\begin{table}[htbp]
\caption{Line coverage by model within each LLM setting. Line~\% is per-app union coverage on the usable line-coverage subset; tokens are per run. The best model in each setting is bold.}
\label{tab:results-setting-model}
\centering
\begin{tabular}{llrr}
\hline
\textbf{Setting} & \textbf{Model} & \textbf{Line \%} & \textbf{Tokens/run} \\
\hline
LLM-guided
  & Claude-Opus-4.8  & 25.2 & 130{,}162 \\
  & Gemini-3.5-Flash & 28.2 & 91{,}149 \\
  & \textbf{GPT-5.5} & \textbf{28.5} & 32{,}967 \\
  & DeepSeek-V4-Pro  & 23.2 & 26{,}521 \\
\hline
LLM-guided-derived
  & Claude-Opus-4.8  & 32.1 & 140{,}445 \\
  & \textbf{Gemini-3.5-Flash} & \textbf{33.7} & 91{,}376 \\
  & GPT-5.5          & 30.7 & 35{,}323 \\
  & DeepSeek-V4-Pro  & 26.1 & 25{,}989 \\
\hline
LLM-generated
  & Claude-Opus-4.8  & 25.8 & 169{,}283 \\
  & \textbf{Gemini-3.5-Flash} & \textbf{29.6} & 102{,}757 \\
  & GPT-5.5          & 27.6 & 52{,}045 \\
  & DeepSeek-V4-Pro  & 22.8 & 13{,}149 \\
\hline
\end{tabular}
\end{table}

\paragraph{Raw crash counts are 82\% noise.} Driving 197 apps produced 1{,}033 crash events, but a content-aware classifier over the post-termination terminal screen, followed by per-case adjudication of every ambiguous exit (Section~\ref{sec:benchmark}), reduces these to \textbf{179 valid TUI-level faults (17.3\%) over 45 apps} (Table~\ref{tab:crash-buckets}).\footnote{Five events remain undecidable as they exited without any error message or traceback.} Noticeable crash sources include apps that printed a usage message and never entered their TUI (156) and graceful Ctrl+C exits (105). The ambiguous middle contained no reservoir of real bugs. Of 233 undecidable exits, adjudication identified only 4 genuine faults and 218 startup/environment failures. \emph{Exit-non-zero is not a crash oracle for this domain.}

\begin{table}[htbp]
\caption{Crash-event triage (1{,}033 events)}
\label{tab:crash-buckets}
\centering
\begin{tabular}{lrr}
\hline
\textbf{Bucket} & \textbf{Events} & \textbf{Share} \\
\hline
\textbf{Valid TUI-level fault} & \textbf{179} & \textbf{17.3\%} \\
Noise (Go cov-emit, usage exit, SIGINT, \dots) & 849 & 82.2\% \\
Undecidable (Non-crash fault) & 5 & 0.5\% \\
\hline
\end{tabular}
\end{table}

\begin{figure}[htbp]
\centering
\includegraphics[width=0.95\columnwidth]{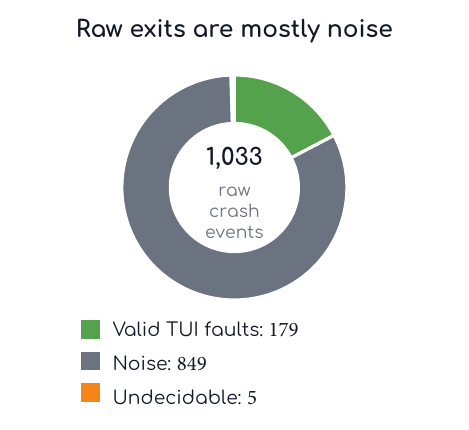}
\caption{Most raw non-zero exits are not TUI faults. The rendered-screen oracle reduces 1{,}033 raw crash events to 179 valid TUI-level faults.}
\label{fig:crash-triage}
\end{figure}

\paragraph{Random wins per run, guidance wins per keystroke.} On valid faults, random has the highest per-run yield (15.2\%),\footnote{Per-run yield is the fraction of runs (one run = one \texttt{(app, setting, model, seed)} session under the fixed 600\,s budget) that surface at least one valid fault; a run that crashes several times still counts once. Random's 15.2\% is 30 of its 197 runs.} ahead of \texttt{LLM-guided-derived} (9.0\%), \texttt{LLM-guided} (8.1\%), and \texttt{LLM-generated} (4.1\%). This lead is, however, a \textbf{pure throughput artifact}. Normalizing by interaction reverses the ranking completely (Table~\ref{tab:crash-efficiency}). Per~1{,}000 steps, the guided LLMs are \textbf{$\sim$13$\times$ more crash-efficient} than random. Random attains the higher per-run yield only because it issues $\sim$24$\times$ more inputs per run. It spends the entire 600\,s emitting keystrokes (median 253 steps), whereas the LLM arms spend most of their wall-clock time awaiting API round-trips and take only a dozen steps. Random is not a better explorer. It is a faster one.

\begin{table}[htbp]
\caption{Valid crashes per run vs.\ per unit of interaction}
\label{tab:crash-efficiency}
\centering
\begin{tabular}{lrrr}
\hline
\textbf{Setting} & \textbf{Med.\ steps} & \textbf{Yield/run} & \textbf{Valid / 1k steps} \\
\hline
random                       & 253 & \textbf{15.2\%} & 0.19 \\
LLM-guided                   & 12  & 8.1\%  & 2.41 \\
\textbf{LLM-guided-derived}  & 13  & 9.0\%  & \textbf{2.57} \\
LLM-generated                & 6   & 4.1\%  & 1.42 \\
\hline
\end{tabular}
\end{table}

In addition, truncating every arm to a matched step budget, at $k\le12$ steps (the LLM arms' median) \texttt{LLM-guided} (5.3\%) and \texttt{LLM-guided-derived} (6.4\%) both exceed random (3.0\%), which overtakes only once granted its full-throughput keystroke budget. This confirms that exploration assisted by the LLM wins per keystroke.

\begin{figure}[htbp]
\centering
\includegraphics[width=\columnwidth]{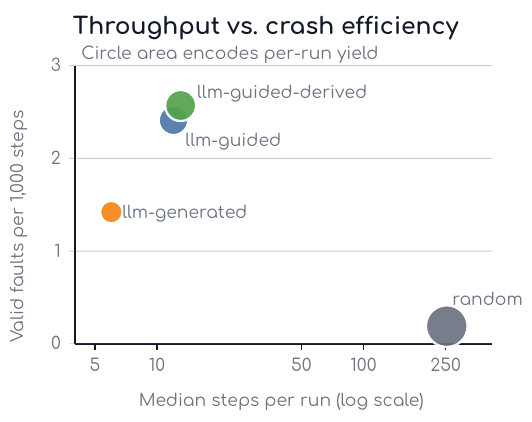}
\caption{The crash result is a throughput--efficiency trade-off. Random issues far more inputs within the fixed time budget, while LLM-guided settings find valid faults much more efficiently per interaction.}
\label{fig:crash-efficiency}
\end{figure}

Three factors explain the picture: (i)~\textbf{Crashes are shallow.} The median valid crash is reached in just 5 steps (115 of 179 within 15 steps), so a few hundred blind keystrokes stumble onto shallow bugs without guidance, and the guidance premium would only show on deep, state-gated bugs, of which this corpus has few. (ii)~\textbf{The strategies are complementary, not ranked.} Some crashing apps are reached \emph{only} by an LLM arm (15 apps; e.g.\ a Rust panic in \texttt{twitch-tui} that fires only after completing an interactive OAuth wizard, and faults in \texttt{posting}, \texttt{frogmouth}, and \texttt{Bagels} that need a filled form or a valid query), while 10 are \emph{random-only} (e.g.\ a \texttt{kmon} segfault reached by brute volume). (iii)~\textbf{LLM attrition shrinks the denominator, not the hit-rate.} 71--83 runs per LLM setting never executed (API failures, timeouts), and \texttt{LLM-generated} quits within 5 steps in half of its non-crash runs (premature ``task done''), so its 4.1\% understates the explorer. Under a fixed \emph{time} budget with shallow targets, random finds the most crashes, but per unit of interaction guidance is an order of magnitude more efficient and uniquely reaches input-gated bugs. The right design is therefore a hybrid that spends LLM calls to unlock state and random throughput to storm it.

\subsection{Fault characteristics}

The 179 valid faults collapse to 47 unique fault signatures over 45 affected applications. \textbf{Python has the highest fault density} (20.8 per 100 runs, vs.\ Rust 6.5, TS 8.6, Go 2.2), largely because Textual renders clean, attributable tracebacks (\texttt{ScreenStackError}, \texttt{MarkupError}, \texttt{AttributeError}). Its failures are \emph{legible}, not necessarily more frequent. \textbf{Rust contributes the most distinct panics} and the only fatal signals (two SIGSEGVs), the highest-severity crashes in the set. Faults cluster on a few primitive triggers, namely Enter~(37), ArrowDown~(14), \texttt{?}~(14), Tab~(11), and Escape~(11), with the help overlay (\texttt{?}) and field submission (Enter) disproportionately fatal. Opening a help pane or committing a field is where unhandled state lives.

\subsection{RQ3: Coverage decouples from fault-finding}

The setting \texttt{LLM-guided} found many crashes yet does not lead on line coverage, and random leads on neither coverage nor per-step crashes yet attains the highest per-run crash yield. Within the usable line-coverage subset, the coverage--fault relationship is therefore essentially flat, contrary to the conventional expectation that more line coverage should imply more faults found. This pattern is consistent with a \textbf{session-truncation confound}, in which a crash terminates the episode and caps the coverage that can accrue afterward, so that the runs which expose faults are penalized on coverage. Together with the observation that valid TUI crashes are shallow (median depth 5 steps) and broadly reachable, this implies that \textbf{line coverage is a poor proxy for crash-finding in this TUI setting}. The result echoes prior cautions that coverage and test effectiveness need not align~\cite{inozemtseva2014coverage,papadakis2018mutants} and motivates a TUI-specific criterion keyed on reachable interactive states rather than executed lines.

\section{Discussion}
\label{sec:discussion}

Terminal user interfaces sit at the heart of widely used software, from coding assistants and system-management tools to database and network clients, yet to the best of our knowledge their automated testing has received almost no systematic study. We therefore take the first step toward closing that gap, and our results point to several directions for TUI testing research. The central lesson is that the testing problem is not only an action-selection problem. It is also a benchmark-design, measurement, launch-configuration, and oracle-design problem.

\textbf{Metric design is an important research topic.} The completed run shows that raw exit counts and raw coverage numbers are easy to misread. Most non-zero exits are not valid TUI faults, and coverage availability differs sharply across runtime instrumentation paths, and even obtaining reliable line coverage for Go took substantial engineering effort. This difficulty is structural: whereas mobile GUI testing targets just two platforms (Android and iOS) with a small set of languages, TUI development is fragmented across many frameworks written in different languages, so a cross-framework study must solve instrumentation and measurement anew for each language rather than once. Future TUI benchmarks should therefore treat measurement validity as a first-class design goal: report valid faults rather than raw exits, and normalize crash-finding by interaction as well as by run.

\textbf{Crash-finding is a tractable first target.} This paper focuses on failures that are visible from the terminal session: tracebacks, panics, fatal signals, and abnormal terminations with rendered evidence. That scope gives the field a concrete starting point. It makes automated triage possible and exposes real faults in widely used applications. The next step is to expand the oracle beyond crashes, toward visual misrenderings, incorrect state transitions, performance stalls, and usability defects. Doing so will require richer ground truth than exit status or stack traces, likely combining replayable interaction logs, screen-state differencing, and application-specific assertions.

\textbf{Hybrid exploration can be a practical path.} Under a fixed wall-clock budget, random exploration remains strong because it emits many more inputs than an LLM-guided policy. At the same time, LLM guidance is much more efficient per interaction and reaches input-gated faults that random input misses. This argues against treating random and LLM-guided testing as mutually exclusive baselines. A practical TUI tester should use LLMs where semantic reasoning matters: deriving launch arguments, creating fixtures, filling forms, and unlocking states. Once the application is in a meaningful state, high-throughput exploration can still be valuable for stressing shallow or brittle transitions.

\textbf{Launch derivation deserves more attention.} A recurring source of low coverage and false crashes is not poor navigation but failure to start the TUI in a meaningful mode. Many applications require files, command-line arguments, credentials, databases, or setup state before their interface appears. The \texttt{llm-guided-derived} setting shows that deriving these launch inputs can matter more than choosing the next keypress. Future work should study launch configuration as its own subproblem, with benchmarks that measure whether a testing system can infer required fixtures and enter the interactive UI before exploration begins.

\textbf{Multimodal models are a natural next step.} Our LLM-guided arms reason from textual representations of the terminal state. Modern multimodal models could instead consume rendered screenshots directly, potentially using layout, styling, alignment, and visual salience that are lost or flattened in a text-grid representation. Screenshot-based policies may be especially useful for TUIs whose state is encoded through color, highlighting, tables, charts, or spatial grouping. They should, however, be evaluated against the same baselines used here: a fixed wall-clock budget, per-interaction normalization, random throughput, and a content-aware crash oracle.

\textbf{The benchmark is a starting point.} The 197 applications in this study form a practical, runnable, cross-framework slice of modern TUIs. They are not a complete census of terminal software. Future benchmark extensions should add other ecosystems, include applications that need controlled external services or accounts, and revive the historical-crash benchmark once buggy versions can be rebuilt and replayed systematically. The current benchmark establishes the measurement foundation on which those broader fault and correctness studies can build.

\section{Threats to Validity}
\label{sec:threats}

\textbf{Construct validity.} Our main construct validity includes coverage, interaction efficiency, and valid TUI-level crashes. Line coverage is an imperfect proxy for TUI behaviour because many terminal-state changes occur through framework rendering code or data-dependent UI state that may not map cleanly to newly executed application lines. Widget coverage is closer to the interface layer but the order of widgets being executed may matter more. Crash validity is also non-trivial: raw non-zero exits include usage errors, graceful interrupts, harness failures, and coverage artifacts. We mitigate this by classifying crashes from the rendered terminal screen and manually adjudicating ambiguous cases, with two of the authors independently labeling each ambiguous event and resolving disagreements by discussion until they reach agreement, but some valid faults may still be missed if they leave no recognizable on-screen evidence.

\textbf{Internal validity.} The fixed wall-clock budget introduces a throughput confound. Random exploration issues hundreds of inputs per run, while LLM-guided exploration often issues only a dozen because of LLM reasoning time. We address this by reporting both per-run yield and valid faults per 1{,}000 steps.

\textbf{External validity.} The benchmark targets four dominant frameworks: ratatui, bubbletea, textual, and ink. Results may not transfer to TUIs built with other frameworks and programming languages, and custom terminal engines. The corpus is also limited to apps that build and run headlessly in Docker. Applications that require live network services, user credentials, hardware, desktop integration, or unusual terminal features are underrepresented and hard to be included.

\section{Related Work}
\label{sec:related}

\textbf{Testing terminal user interfaces.} Despite the prevalence of TUI applications, automated testing of the terminal interface itself has received almost no systematic study. The testing support that exists is framework-native and practitioner-built rather than the subject of published research. Each major framework ships a unit-level harness that renders the interface into an in-memory buffer and compares it against a stored frame, including the ratatui \texttt{TestBackend}~\cite{ratatui}, the bubbletea \texttt{teatest} model harness~\cite{bubbletea}, the textual \texttt{run\_test} Pilot API~\cite{textual}, and the \texttt{ink-testing-library} \texttt{lastFrame} probe~\cite{ink}. End-to-end drivers that operate a real pseudo-terminal exist at the tooling level~\cite{tuitest}. These harnesses share two limitations that our study documents quantitatively. They are exercised by hand-written tests whose coverage of the interface is shallow (Section~\ref{sec:motivation}), and they assume a clean process exit, which an interactive application under exploration does not provide (Section~\ref{sec:benchmark}). No prior work establishes a cross-framework benchmark, an instrumentation method that recovers coverage from a terminated session, or an empirical account of how automated techniques perform on TUIs. This paper supplies all three, and in doing so treats TUI testing as a research problem in its own right rather than a downstream application of GUI or CLI methods.

\textbf{Automated GUI exploration.} The body of work nearest to our exploration component is automated testing of graphical applications, where a driver issues input events and observes the resulting interface state. Random and model-free strategies set the baseline. Dynodroid generates input events through a guided observe-select-execute loop~\cite{machiry2013dynodroid}, and the random ``monkey'' tester remains the reference point that more elaborate techniques are measured against. Choudhary et al.\ ran exactly that comparison and found that none of the surveyed academic tools reliably beat random exploration on coverage~\cite{choudhary2015monkey}, a result that anticipates our own. Later work moved to model-based and search-based drivers, with Stoat building a stochastic model of the interface~\cite{su2017stoat}, Sapienz casting test generation as multi-objective search~\cite{mao2016sapienz}, and reinforcement learning treating exploration as a sequential decision problem~\cite{romdhana2022drl}. Every one of these techniques relies on the addressability a GUI provides, namely a widget tree with stable identifiers, roles, and bounding boxes against which an action can be aimed and a result queried. A TUI exposes none of this. Its observable state is an $H\times W$ grid of styled character cells, and its input is a raw byte stream, so the perception and action interfaces these methods assume do not exist. We retain the random baseline these papers established and ask, for the first time, how it and its learned competitors behave when the only observation is a rendered screen.

\textbf{LLM-driven testing.} A second line of work replaces hand-built exploration policies with large language models. For mobile GUIs, Liu et al.\ prompt an LLM to choose functionality-aware actions during testing~\cite{liu2024guiGPT} and to synthesise unusual text inputs that trigger crashes~\cite{liu2024unusualinputs}. For library and unit testing, LLMs have been used as zero-shot fuzzers of deep-learning frameworks~\cite{deng2023titanfuzz} and as generators of unit test suites~\cite{schafer2024llmunit}. These studies report that LLM guidance improves over weaker baselines on their respective targets, but each operates where the model can read structured artefacts such as an accessibility tree, a method signature, or library documentation. None addresses an application whose entire interface is a character grid driven by raw key events, and none reports the throughput cost that a per-step LLM call imposes under a fixed wall-clock budget. We evaluate LLM-guided exploration, LLM-derived launch inputs, and LLM-generated test scenarios on TUIs under a time-equalised budget, and we find that the per-keystroke advantage of guidance is real but is repeatedly outrun by the sheer throughput of random input, a trade-off prior LLM-testing work does not surface.

\textbf{Coverage and fault-finding.} Whether code coverage predicts a test suite's ability to find faults is a long-running question, and the evidence is mixed even within a single domain. Inozemtseva and Holmes report that coverage is not strongly correlated with suite effectiveness once suite size is controlled~\cite{inozemtseva2014coverage}, and Papadakis et al. qualify the use of mutants as fault proxies at scale~\cite{papadakis2018mutants}. In GUI testing specifically, coverage and fault detection have been found to move together more often~\cite{strecker2008relationships,yuan2011gui}, and coverage continues to serve as the default proxy in fuzzer benchmarking~\cite{bohme2022fuzzbench} and fault localisation~\cite{silva2023flacoco}. Our results place TUIs on the sceptical side of this debate and explain why. A crash truncates the exploration session that found it, capping the coverage that run can accrue, so the very executions that expose faults are penalised on coverage. The decoupling we observe is therefore not noise but a structural property of interactive, crash-terminated testing, which is precisely the regime a TUI tester operates in, and it motivates a coverage criterion keyed on reachable interactive states rather than executed lines.

\textbf{Empirical ``how far are we'' studies.} Our paper follows the empirical tradition that measures the real gap between a technique's promise and its delivered behaviour rather than proposing a new tool. Choudhary et al. asked whether automated Android input generation had arrived and answered with a sober comparison against random~\cite{choudhary2015monkey}; Lukasczyk et al. subjected automated unit-test generation for Python to the same scrutiny~\cite{lukasczyk2023pynguin}; and Tian et al. assessed how far a frontier LLM actually goes as a programming assistant~\cite{tian2023chatGPT}. Each of these studies derives its value from honest baselines, careful confound analysis, and an open artefact, and recent guidance for LLM-based software-engineering experiments codifies exactly these practices~\cite{baltes2025guidelines}. No such study exists for TUIs, because there has been no benchmark on which to run one. We provide that benchmark, hold the LLM arms to a model-free random baseline and a content-aware crash oracle, and report where automated TUI testing stands today.

\section{Conclusion}
\label{sec:conclusion}

Terminal User Interfaces are a large and growing class of software that is poorly tested in practice and lacks a dedicated testing methodology. We presented the first open, multi-language empirical study of automated TUI testing, over a benchmark of 197 headlessly-runnable, coverage-instrumented applications across the four dominant frameworks, driven by four frontier LLMs and a random baseline under an equalized wall-clock budget.

No single model dominates, and capability is decoupled from cost. The largest gain comes from automatically deriving launch inputs, not from smarter per-step reasoning; and while random looks competitive under a time budget, per interaction LLM guidance is an order of magnitude more efficient and uniquely reaches input-gated faults, arguing for hybrid strategies. Measurement is itself a research problem: raw exits are 82\% noise, and once an oracle isolates the 179 valid faults, line coverage proves a poor proxy for crash-finding. Automated TUI testing is feasible but far from solved, and we hope this study establishes it as a research area in its own right. We release the cross-language coverage tool \texttt{tuicov} at \url{https://github.com/tui-testing/tuicov} and the automated TUI testing framework \texttt{tuibot} at \url{https://github.com/tui-testing/tuibot}.

\bibliographystyle{plainnat}
\bibliography{references}

\end{document}